\documentclass[showpacs,showkeys,preprintnumbers]{revtex4}%
\usepackage{amsfonts}
\usepackage{amsmath}
\usepackage{graphicx}
\usepackage{dcolumn}
\usepackage{bm}
\usepackage{amssymb}%
\begin{document}
\title{Energy uncertainty of the final state of a decay process }
\author{Francesco Giacosa}
\affiliation{\emph{Institut f\"{u}r Theoretische Physik Johann Wolfgang Goethe -
Universit\"{a}t, Max von Laue--Str. 1 D-60438 Frankfurt, Germany}}

\begin{abstract}
We derive the expression for the energy uncertainty of the final state of a
decay of an unstable quantum state prepared at the initial time $t=0$. This
expression is function of the time $t$ at which a measurement is performed to
determine if the state has decayed and, if yes, in which one of the infinitely
many possible final states. For large times the energy spread is, as expected,
given by the decay width $\Gamma$ of the initial unstable state. However, if
the measurement of the final state is performed at a time $t$ comparable to
(or smaller than) the mean lifetime of the state $1/\Gamma$, then the
uncertainty on the energy of the final state is much larger than the decay
width $\Gamma$. Namely, for short times an uncertainty of the type $1/t$
dominates, while at large times the usual spread $\Gamma$ is recovered. Then,
we turn to a generic two-body decay process and describe the energy
uncertainty of each one of the two outgoing particles. We apply these formulas
to the two-body decays of the neutral and charged pions and to the spontaneous
emission process of an excited atom. As a last step, we study a case in which
the non-exponential decay is realized ad show that for short times eventual
asymmetric terms are enhanced in the spectrum.

\end{abstract}

\maketitle

It is well known that an unstable quantum state, which we denote as
$\left\vert S\right\rangle $, is not an energy eigenstate of the full
Hamiltonian $H$ of the corresponding quantum system, but shows an energy
spread of the order of its decay width $\Gamma$. One defines for such an
unstable state a `spectral function' $d_{S}(E)$, which represents the
probability distribution of energy (thus implying the validity of the
normalization $\int_{-\infty}^{+\infty}d_{S}(E)dE=1$). The survival
probability amplitude, i.e. the amplitude that the unstable state $\left\vert
S\right\rangle $ prepared at the time $t=0$ has not yet decayed at the later
instant $t,$ is given by the Fourier transform of the spectral function
\cite{khalfin,ghirardi},%
\begin{equation}
a(t)=\left\langle S\right\vert e^{-iHt}\left\vert S\right\rangle
=\int_{-\infty}^{+\infty}d_{S}(E)e^{-iEt}dE\text{ ;} \label{at}%
\end{equation}
the survival probability $p(t)$ is given as the square of the amplitude,
$p(t)=\left\vert a(t)\right\vert ^{2}$. (Note: natural units $c=\hslash=1$ are used.)

In the very well known, and in most cases very well accurate, Breit-Wigner
(BW) limit \cite{ww,breit}, the spectral function takes the Lorentzian form
\begin{equation}
d_{S}(E)\overset{\text{BW-limit}}{=}\frac{\Gamma}{2\pi}\frac{1}{(E-M)^{2}%
+\Gamma^{2}/4}, \label{bwdse}%
\end{equation}
where $M$ is referred to as the `mass' (or energy) of the unstable state
$\left\vert S\right\rangle $. In this limit the exponential form
\begin{equation}
a(t)\overset{\text{BW-limit}}{=}e^{-iMt-\Gamma t/2}\rightarrow p(t)=e^{-\Gamma
t}. \label{abw}%
\end{equation}
is obtained by using Eq. (\ref{at}). It is a fact that the BW form is not
exact; the existence of an energy threshold, $d_{S}(E)=0$ for $E<E_{th}$, is
necessary for the quantum system to be consistent; in turn, this property
implies that $p(t)$ is not exponential for long times \cite{khalfin,ghirardi},
see also the discussions in Ref. \cite{mercouris} and refs. therein; for an
indirect experimental proof of long-time deviations we refer to
\cite{berillium} and for a direct one which makes use of decays of organic
materials to \cite{rothe}. Deviations from the BW form for large values of the
energy ($E\gg M$), which usually take place due to a form factor that makes
the function $d_{S}(E)$ decreasing faster than $E^{-2}$ for large values of
$E$, implies a non-exponential behavior of $p(t)$ at short times. These
short-time deviations have been confirmed experimentally \cite{raizen1}, which
in turn have also led to the verification of the Zeno and Anti-Zeno effects
\cite{raizen2}. (For theoretical details in Quantum Mechanics see Refs.
\cite{ghirardi,misra,facchiprl,duecan} and in the framework of Quantum Field
Theory (QFT) Refs. \cite{duecan,zenoqft,pascazioqft}; for a general discussion
of spectral functions and deviations from the BW limit in QFT see also Refs.
\cite{salam,achasov,lupo,lupo2,e38}.)

Quite recently, a vivid debate on the non-exponential weak decay of ions
measured in\ Ref. \cite{gsianomaly} has taken place. This is an extremely
interesting phenomenon because for the first time short-time deviations from
the exponential decay have been seen in a microscopic and natural nuclear
system. In Ref. \cite{gp} these deviations have been linked to a modification
of the BW distribution due to interactions with the measuring apparatus. Other
explanations, based on neutrino oscillations and energy level splitting of the
initial state, have been put forward \cite{giunti}. At present, no consensus
on the measured data exist.

Often the theoretical interest in the study of a decay has focussed on the
determination of the survival probability $p(t)$ of the initial unstable state
$\left\vert S\right\rangle $. In this work we turn our attention to the final
states of the decay of the unstable state. To this end, we suppose to perform
at the instant $t>0$ a measurement on the quantum system of the following
type: we measure the probability that the quantum state has decayed and, if
yes, in which one of its possible final states. In fact, an infinity of such
finals states is present in each decay process; at a given instant $t$, the
sum of all these probabilities must clearly be the decay probability $1-p(t).$
Still, the question of the value of each single probability of decay in a
given final state (and not only the overall sum), is interesting and sheds
light on the distribution of energy of the final state in a decay process.

In order to make our discussion quantitative, we need to fix an Hamiltonian.
We use the quite general Lee Hamiltonian approach \cite{lee,chiu}, in which
the unstable state $\left\vert S\right\rangle $ is coupled to an infinity set
of final states $\left\vert k\right\rangle $:
\begin{equation}
H=H_{0}+H_{1}\text{ ,}%
\end{equation}
where the free Hamiltonian $H_{0}$ and the interacting Hamiltonian $H_{1}$ are
given by
\begin{equation}
H_{0}=M\left\vert S\right\rangle \left\langle S\right\vert +\int_{-\infty
}^{+\infty}dk\omega(k)\left\vert k\right\rangle \left\langle k\right\vert
\text{ , }H_{1}=\int_{-\infty}^{+\infty}\frac{dk}{\sqrt{2\pi}}gf(k)\left(
\left\vert k\right\rangle \left\langle S\right\vert +\left\vert S\right\rangle
\left\langle k\right\vert \right)  \text{ .}%
\end{equation}
The quantity $g$ is a coupling constant with the dimension of energy$^{1/2}$.
The dimensionless function $f(k)$ specifies the mixing of the state
$\left\vert S\right\rangle $ with the state $\left\vert k\right\rangle ;$ the
energy $\omega(k)$ is the energy of the state $\left\vert k\right\rangle $ in
the interaction free case. Moreover, the following normalizations hold:
$\left\langle S|S\right\rangle =1,$ $\left\langle k|k^{\prime}\right\rangle
=\delta(k-k^{\prime})$, $\left\langle S|k\right\rangle =0,$ $1=\left\vert
S\right\rangle \left\langle S\right\vert +\int_{-\infty}^{+\infty}dk\left\vert
k\right\rangle \left\langle k\right\vert $. Note, we have taken for simplicity
$k$ as a one-dimensional variable: if we think of a two-body decay of the
state $\left\vert S\right\rangle $ in its rest frame, the ket $\left\vert
k\right\rangle $ describes a two-particle state, one of which is moving with
momentum $k$ and the other one with momentum $-k.$ The generalization to a
three-dimensional decay $\vec{k}$ is straightforward, but unnecessary for our
purposes, see Ref. \cite{duecan} which we also refer to for further technical details.

At the initial time $t=0$ the state $\left\vert S\right\rangle $ is prepared.
Then, the time evolution implies that at the instant $t$ the system is
described by the state%
\begin{equation}
e^{-iHt}\left\vert S\right\rangle =a(t)\left\vert S\right\rangle
+\int_{-\infty}^{+\infty}\frac{dk}{\sqrt{2\pi}}a_{Sk}(t)\left\vert
k\right\rangle \text{ .} \label{te}%
\end{equation}
A possible way to evaluate the previous expression makes use of the operator
relation $e^{-iHt}=\frac{i}{2\pi}\int_{-\infty}^{+\infty}dEG(E)e^{-iEt}$ with
$G(E)=\left[  E-H+i\varepsilon\right]  ^{-1}$ \cite{facchiprl,fpbook}. Then,
the validity of Eq. (\ref{at}) can be proven:
\begin{equation}
a(t)=\frac{i}{2\pi}\int_{-\infty}^{+\infty}dEG_{S}(E)e^{-iEt}\text{ }%
=\int_{-\infty}^{+\infty}dEd_{S}(E)e^{-iEt}\text{ ,}%
\end{equation}
where the propagator $G_{S}(E)$ of the unstable state $\left\vert
S\right\rangle $ reads
\begin{equation}
G_{S}(E)=\left\langle S\left\vert G(E)\right\vert S\right\rangle =\frac
{1}{E-M+\Pi(E)+i\varepsilon}\text{ , }\Pi(E)=\int_{-\infty}^{+\infty}\frac
{dk}{2\pi}\frac{g^{2}f^{2}(k)}{E-\omega(k)+i\varepsilon}\text{ ,} \label{gse}%
\end{equation}
and the spectral function emerges as the imaginary part of the propagator:
$d_{S}(E)=\frac{1}{\pi}\left\vert \operatorname{Im}G_{S}(E)\right\vert .$ The
quantity $\Pi(E)$ is referred to as the self-energy quantum contribution for
the state $\left\vert S\right\rangle $.

The BW limit is obtained for $f(k)=1$ and $\omega(k)=k$, that implies
$\Pi(E)=i\Gamma/2$, where the decay width $\Gamma$ reads: $\Gamma=g^{2}$. In
this limit Eq. (\ref{abw}) follows. For generic $f(k)$ and $\omega(k)$, the
decay is not exponential, but is usually very well approximated by an
exponential decay where the decay width is given by the Fermi golden-rule:
$\Gamma=2\operatorname{Im}[\Pi(M)]=g^{2}f^{2}(k_{M})/\left\vert \omega
^{\prime}(k_{M})\right\vert $, where $\omega(k_{M})=M$. Note, the choice
$\omega(k)=k$ means that the energy is not bounded from below. This is
obviously unphysical and represents a mathematical trick. However, as long as
the distribution function is peaked around $M$ and the low-energy threshold is
far away from it, the error done by this approximation is (indeed very) small.

In this work we are interested in the probability to find the system described
by a certain state $\left\vert k\right\rangle $ when performing a measurement
at the instant $t.$ To this end, it is important to determine the coefficients
$a_{Sk}(t)$ in Eq. (\ref{te}). A direct evaluation delivers the following
general result:%
\begin{equation}
a_{Sk}(t)=\left\langle k\right\vert e^{-iHt}\left\vert S\right\rangle
=i\frac{gf(k)}{\sqrt{2\pi}}\int_{-\infty}^{+\infty}\frac{dk}{2\pi}\frac
{G_{S}(E)}{E-\omega(k)+i\varepsilon}e^{-iEt}\text{ .} \label{askgen}%
\end{equation}
It follows that the decay probability, i.e. the probability that the state has
decayed at the time $t,$ is given by \textquotedblleft one minus the survival
probability\textquotedblright:%
\begin{align}
w(t)  &  =\int_{-\infty}^{+\infty}dk\left\vert a_{Sk}(t)\right\vert
^{2}=\text{ }\int_{-\infty}^{+\infty}dk\left\langle S\right\vert
e^{iHt}\left\vert k\right\rangle \left\langle k\right\vert e^{-iHt}\left\vert
S\right\rangle \nonumber\\
&  =\left\langle S\right\vert e^{iHt}\left(  1-\left\vert S\right\rangle
\left\langle S\right\vert \right)  e^{-iHt}\left\vert S\right\rangle
=1-p(t)\text{ } \label{w}%
\end{align}
In addition to this, the previous expression also tells us that the
probability density that a particular final state $\left\vert k\right\rangle $
is realized at the instant $t$ is given by the quantity $\left\vert
a_{Sk}(t)\right\vert ^{2}$. By a change of coordinate from $k$ to
$\omega=\omega(k)$ we rewrite $w(t)$ as%
\begin{equation}
w(t)=\int_{-\infty}^{+\infty}d\omega\eta(t,\omega)\text{ ,} \label{etaint}%
\end{equation}
where the quantity $\eta(t,\omega)d\omega$ is the probability that, by a
measurement of the final state of the quantum state at the instant $t$, the
quantum state has decayed and has an energy between $\omega$ and
$\omega+d\omega$.

Let us in the following restrict to the BW limit, which, as mentioned above,
is realized for $f(k)=1$ and $\omega(k)=k$. In this case a simple explicit
calculation shows that the energy distribution $\eta(t,\omega)$ takes the
form
\begin{equation}
\eta(t,\omega)=\frac{\Gamma}{2\pi}\left\vert \frac{e^{-i\omega t}%
-e^{-i(M-i\Gamma/2)t}}{\omega-M+i\Gamma/2}\right\vert ^{2}. \label{eta}%
\end{equation}
A direct evaluation of the integral (\ref{etaint}) using the explicit BW form
of Eq. (\ref{eta}) shows the validity of the general Eq. (\ref{w}) in the BW
limit: $w(t)=\int_{-\infty}^{+\infty}d\omega\eta(t,\omega)=1-e^{-\Gamma t}$.

The validity of $\eta(t,\omega)$ in Eq. (\ref{eta}) is general, as long as the
exponential (or BW) limit is considered. It depends only on the two parameters
$M$ and $\Gamma$ and not on the details of the decay process under study. In
particular, Eq. (\ref{eta}) does not depend on the fact that our variable $k$
is one dimensional; the same result would be obtained with a more realistic
three-dimensional variable. The result also does not depend on the special
form of the Lee-Hamiltonian, but would be obtained with each Hamiltonian, as
long as the BW limit is taken. Quite remarkably, the result is applicable also
in relativistic Quantum\ Field Theory because the nonrelativistic BW limit is
typically a very good approximation also in this context, see details in Ref.
\cite{duecan} in which the link between QFT and Lee models has been described.
Clearly, when including deviations from the Breit-Wigner energy distribution,
also $\eta(t,\omega)$ changes (see the end of the paper), but, as long as the
exponential decay law well approximates the decay of an unstable state, Eq.
(\ref{eta}) represents a very good description of the probability density of
the energy of the final state for a wide range of $\omega$.%

\begin{figure}
[ptb]
\begin{center}
\includegraphics[
height=3.8337in,
width=5.5988in
]%
{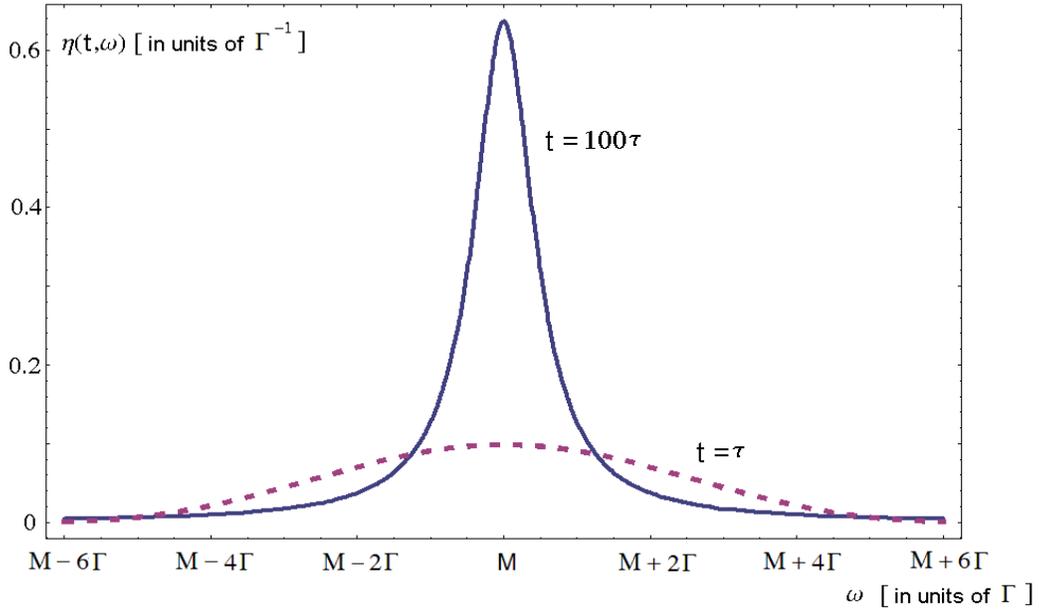}%
\caption{(Color online) The functions $\eta(100\tau,\omega)$ (solid line) and
$\eta(\tau,\omega)$ (dashed line) defined in Eq. (\ref{eta}) are plotted as
function of the energy $\omega.$ The quantity $\eta(100\tau,\omega)$ is to a
very good extent the BW distribution of Eq. (\ref{bwdse}), while $\eta
(\tau,\omega)$ is sizably wider.}%
\end{center}
\end{figure}

The energy distribution of Eq. (\ref{eta}) was already implicitly present in
the pioneering work of Ref. \cite{ww} where the natural broadening of spectral
lines was studied, see also e.g. Refs.
\cite{scully,facchispont,ford,elattari,kofman} and refs. therein. In
particular, in\ the limit of large times, $t\rightarrow\infty$, one obtains:
\begin{equation}
\eta(t\rightarrow\infty,\omega)=\frac{\Gamma}{2\pi}\left\vert \frac{1}%
{\omega-M+i\Gamma/2}\right\vert ^{2}=\frac{\Gamma}{2\pi}\frac{1}%
{(\omega-M)^{2}+\Gamma^{2}/4}\text{ .}%
\end{equation}
Thus, in the long-time regime the energy distribution is the original energy
distribution of the initial state (\ref{bwdse}). The probability that the
final state has an energy between $\omega$ and $\omega+d\omega$ is given by
$\eta(t\rightarrow\infty,\omega)d\omega=d_{S}(\omega)d\omega$. This result has
been shown in Ref. \cite{ford} and can be also easily derived with the general
expressions written above; this is indeed a general outcome which is not
restricted to the BW approximation.

In the present work we aim to discuss the distribution of energies of the
final state also for early times and not only in the long-time limit. Namely,
we study the form of $\eta(t,\omega)$ of Eq. (\ref{eta}) as function of
$\omega$ for different values of $t.$ The function has, for each value of $t$,
a maximum for $\omega=M$. The value of the function at its maximum
\begin{equation}
\eta(t,\omega=M)=\frac{2}{\pi\Gamma}\left(  1-e^{-\Gamma t}\right)  ^{2}%
\end{equation}
vanishes for $t\rightarrow0$ and increases for increasing $t$; this is in
agreement with the fact that the overall area of $\eta(t,\omega)$ is a
increasing function of $t.$ In Fig. 1 we plot the function $\eta(t,\omega)$
for $t=\tau$ and for $t=100\tau$ where $\tau=\Gamma^{-1}$ is the mean
lifetime. It is visible that for $t=100\tau$ the limit $t\rightarrow\infty$ is
well recovered. However, for $t=\tau$ the function $\eta(t=\tau,\omega)$ is
sizably wider, implying a broadening of the energy uncertainty if a
measurement of the final state (and its energy) is performed at such an early time.

In Fig. 2 we plot the function $\eta(t,\omega)/\eta(t,M)$ for $t=\tau/10,$
$\tau/2,\tau,3\tau,100\tau$. Being this quantity per construction normalized
to 1 at the maximum $\omega=M$, the spread of the energy distribution of the
final state is better visualized. For short times, $t\lesssim$ $3\tau$, the
function $\eta(t,\omega)$ is spread over a wide range of energy, sizably
larger than the natural expected energy spread $\Gamma.$ Moreover, a pattern
of maxima and minima become visible for small values of $t$.%

\begin{figure}
[ptb]
\begin{center}
\includegraphics[
height=3.7766in,
width=5.7527in
]%
{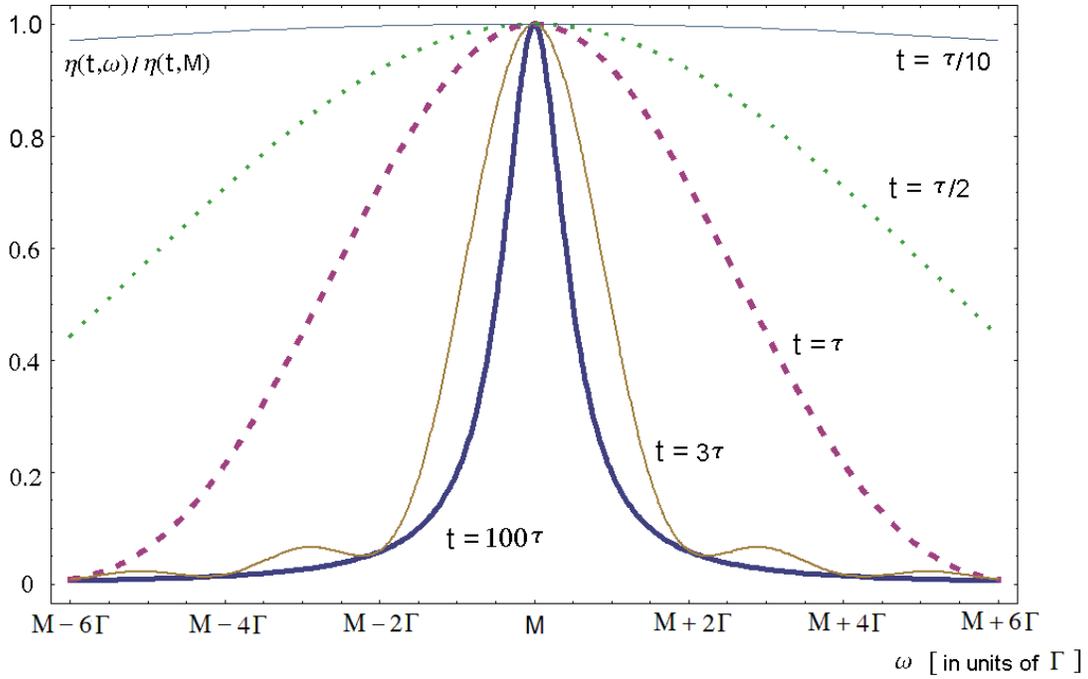}%
\caption{(Color online) The function $\eta(t,\omega)/\eta(t,M),$ which is
normalized to $1$ for $\omega=M,$ is plotted as function of $\omega$ for
$t=\tau/10$ (solid thin line) $\tau/2$ (dotted line), $\tau$ (dashed thick
line), $3\tau$ (solid thin line), and $100\tau$ (solid thick line). The
increasing width of the function for smaller times is clearly visible.}%
\end{center}
\end{figure}

In order to quantify the energy uncertainty of the final state, the variance
is not usable because the integral $\int_{-\infty}^{\infty}\omega^{2}%
\eta(t,\omega)d\omega$ is divergent in the BW limit (more in general, it is
dependent on the tails of the distribution). A well-defined quantity for our
purposes is the width, at a given time $t$, of the distribution $\eta
(t,\omega)$ at mid height. We denote it as $\delta\omega=\delta\omega(t)$,
mathematically expressed by the equation
\begin{equation}
\frac{\eta(t,M)}{2}=\eta(t,M+\delta\omega/2)\text{ .} \label{deltaomega}%
\end{equation}
The function $\delta\omega/\Gamma$ is plotted in Fig. 3 as function of $t$:
for large $t$ the quantity $\delta\omega/\Gamma$ tends to $1,$ which is the
`natural' energy uncertainty of the initial state. Conversely, for short times
$\delta\omega/\Gamma$ increases sizably. The following approximate form is
valid for $t\lesssim3\tau$
\begin{equation}
\delta\omega\simeq\frac{2\cdot2.78}{t}\text{ .} \label{deltapprx}%
\end{equation}
The numerical coefficient $2.78$ is the (nonzero) solution of the
transcendental equation $y=\sqrt{2}\left\vert 1-e^{iy}\right\vert $, which
follows from Eq. (\ref{deltaomega}) in the short-time limit. Eq.
(\ref{deltapprx}) approximates the form of $\delta\omega$ better and better
for decreasing time, as it is shown in Fig. 3. This fact is intuitively
expected: for short times the energy uncertainty becomes dominant and this
uncertainty is proportional to $1/t$. The full form of $\delta\omega$
interpolates between the $1/t$ behavior at short times and the constant limit
$\Gamma$ for long times.

In Ref. \cite{elattari} a related phenomenon has been studied in the case of
electron tunneling out of a quantum dot. The attention is focussed on the
continuous monitoring of the unstable state: it is shown that an increase of
the width of the emitted spectrum arises. This is similar to our result,
although the physical situation has some important differences. Namely, in
Ref. \cite{elattari} the continuous measurement (modelled by a term in the
Hamiltonian) is acting on the unstable initial state ($\left\vert
S\right\rangle $ in our notation), while in our framework the (instantaneous)
measurement is designed to detect the final decay product (one of the states
$\left\vert k\right\rangle $). The broadening of our study is an intrinsic
feature of the short-time evolution and is not associated to the measurement
process, which is of the ideal type.

In Ref. \cite{kofman} a formalism which is similar to the one used in this
work has been adopted to study atomic spontaneous emission. Different choices
for the interaction of the unstable upper level to the ground state and to an
emitted photon are extensively analyzed. We thus refer to this work for
further important details and for the investigation of non-exponential decay
law. However, in Ref. \cite{kofman} the emitted spectrum is investigated only
in the limit $t\rightarrow\infty$ and not for earlier times. As we shall see
later on, the spontaneous emission of an atom can be seen as a particular
limit of our approach.

%

\begin{figure}
[ptb]
\begin{center}
\includegraphics[
height=3.5405in,
width=5.559in
]%
{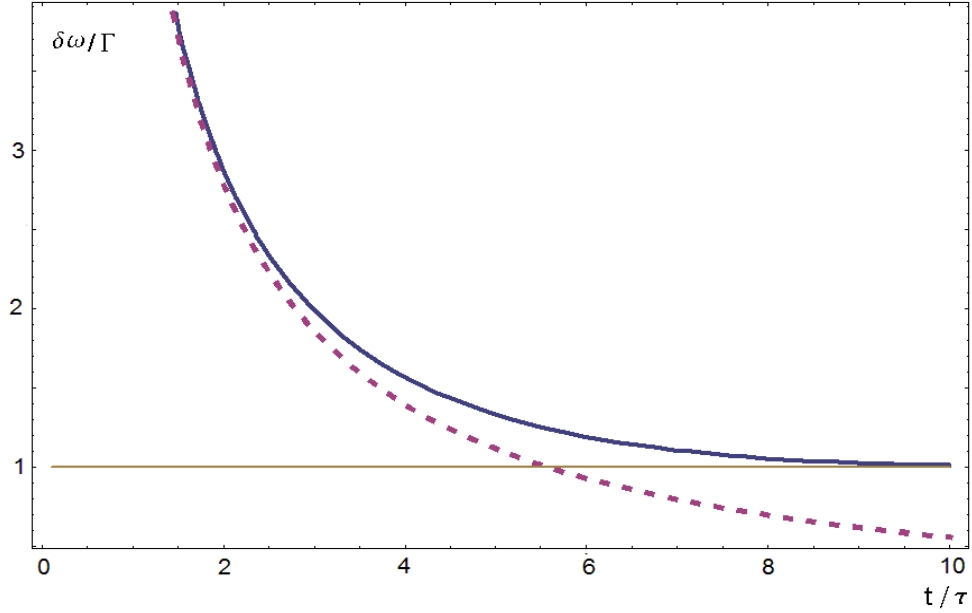}%
\caption{(Color online) The quantity $\delta\omega/\Gamma$ defined in Eq.
(\ref{deltaomega}) is plotted as function of $t/\tau$ (solid line): for short
times it increases as $5.56/t$ (dashed line), while for large times it reaches
the unity, being -as expected- $\delta\omega=\Gamma$ for $t\rightarrow\infty
$.}%
\end{center}
\end{figure}

A nice feature of the Lee Hamiltonian used in this work is that the formalism
can be easily applied to the general case of a two-body decay, which we
describe in the following: the final state $\left\vert k\right\rangle $ with
energy $\omega$ describes two particles, with masses $m_{1}$ and $m_{2}$
respectively, flying back-to-back. The energies $\omega_{1}$ and $\omega_{2}$
of the first and the second particle read (in a general relativistic
framework):%
\begin{equation}
\omega_{1}=\frac{\omega^{2}+m_{1}^{2}-m_{2}^{2}}{2\omega}\text{ , }\omega
_{2}=\frac{\omega^{2}-m_{1}^{2}+m_{2}^{2}}{2\omega}\text{ .}%
\end{equation}
The constraint $\omega_{1}+\omega_{2}=\omega$ is fulfilled and, for each value
of $\omega$, the energies $\omega_{1}$ and $\omega_{2}$ are uniquely
determined. The opposite is not true, being $\omega=\omega_{1}\pm\sqrt
{\omega_{1}^{2}-m_{1}^{2}+m_{2}^{2}}$ and $\omega=\omega_{2}\pm\sqrt
{\omega_{2}^{2}+m_{1}^{2}-m_{2}^{2}}$. One can determine in a straightforward
way also the general expression for the probability density distribution
$\eta_{1}(t,\omega_{1})$ ($\eta_{2}(t,\omega_{2})$), which represents the
probability that the first (second) particle has an energy between $\omega
_{1}+d\omega_{1}$ ($\omega_{2}+d\omega_{2}$) if a measurement at the instant
$t$ on it is performed:%
\begin{equation}
\eta_{1}(t,\omega_{1})=\eta_{1}^{+}(t,\omega_{1})+\eta_{1}^{-}(t,\omega
_{1})\text{ , }\eta_{1}^{\pm}(t,\omega_{1})=\left\vert 1\pm\frac{\omega_{1}%
}{\sqrt{\omega_{1}^{2}-\Delta m^{2}}}\right\vert \eta\left(  t,\omega_{1}%
\pm\sqrt{\omega_{1}^{2}-\Delta m^{2}}\right)  \label{eta1}%
\end{equation}%
\begin{equation}
\eta_{2}(t,\omega_{2})=\eta_{2}^{+}(t,\omega_{2})+\eta_{2}^{-}(t,\omega
_{2})\text{ , }\eta_{2}^{\pm}(t,\omega_{2})=\left\vert 1\pm\frac{\omega_{2}%
}{\sqrt{\omega_{2}^{2}+\Delta m^{2}}}\right\vert \eta\left(  t,\omega_{2}%
\pm\sqrt{\omega_{2}^{2}+\Delta m^{2}}\right)  \text{ ,} \label{eta2}%
\end{equation}
where $\Delta m^{2}=m_{1}^{2}-m_{2}^{2}.$ Notice that, if we choose $\Delta
m^{2}>0$, the function $\eta_{1}(t,\omega_{1})$ is only defined for
$\left\vert \omega_{1}\right\vert \geq\Delta m^{2}$. This limit is however
irrelevant in practical cases because the relation $\omega_{1}>m_{1}$ holds
(relativistically: $\omega_{1}=\sqrt{k^{2}+m_{1}^{2}}$ ). Eqs. (\ref{eta1})
and (\ref{eta2}) are of general validity and can be applied beyond the BW limit.

In most practical cases, when the peak of $d_{S}(E)$ is narrow and away from
threshold, the relation between full energy $\omega$ and the energies of the
two outgoing particles $\omega_{1}$ and $\omega_{2}$ can be simplified:
\begin{equation}
\omega-M\simeq\frac{2M^{2}}{M^{2}-\Delta m^{2}}\left(  \omega_{1}-\bar{\omega
}_{1}\right)  =\frac{2M^{2}}{M^{2}+\Delta m^{2}}\left(  \omega_{2}-\bar
{\omega}_{2}\right)  \text{ , }%
\end{equation}
where
\begin{equation}
\bar{\omega}_{1}=\frac{M^{2}+\Delta m^{2}}{2M}\text{ , }\bar{\omega}_{2}%
=\frac{M^{2}-\Delta m^{2}}{2M}\text{ .} \label{o12b}%
\end{equation}
Introducing the `decay widths'
\begin{equation}
\Gamma_{1}=\left(  \frac{1}{2}-\frac{\Delta m^{2}}{2M^{2}}\right)
\Gamma\text{ , }\Gamma_{2}=\left(  \frac{1}{2}+\frac{\Delta m^{2}}{2M^{2}%
}\right)  \Gamma\text{ ,} \label{gamma12}%
\end{equation}
the expression of the probability distributions $\eta_{1}(t,\omega_{1})$ and
$\eta_{2}(t,\omega_{1})$ takes the simplified form%
\begin{equation}
\eta_{1}(t,\omega_{1})=\frac{2M^{2}}{M^{2}-\Delta m^{2}}\eta\left(
t,M+\frac{2M^{2}\left(  \omega_{1}-\bar{\omega}_{1}\right)  }{M^{2}-\Delta
m^{2}}\right)  =\frac{\Gamma_{1}}{2\pi}\left\vert \frac{1-e^{i\frac
{2M^{2}(\omega_{1}-\bar{\omega}_{1}+i\Gamma_{1}/2)t}{M^{2}-\Delta m^{2}}}%
}{\omega_{1}-\bar{\omega}_{1}+i\Gamma_{1}/2}\right\vert ^{2}\text{ ,}%
\end{equation}%
\begin{equation}
\eta_{2}(t,\omega_{2})=\frac{2M^{2}}{M^{2}+\Delta m^{2}}\eta\left(
t,M+\frac{2M^{2}\left(  \omega_{2}-\bar{\omega}_{2}\right)  }{M^{2}+\Delta
m^{2}}\right)  =\frac{\Gamma_{2}}{2\pi}\left\vert \frac{1-e^{i\frac
{2M^{2}(\omega_{2}-\bar{\omega}_{2}+i\Gamma_{2}/2)t}{M^{2}+\Delta m^{2}}}%
}{\omega_{2}-\bar{\omega}_{2}+i\Gamma_{2}/2}\right\vert ^{2}\text{ ,}%
\end{equation}
where in the r.h.s. of the latter equations the BW-limit has been again taken.
The previous study of the function $\eta(t,\omega)$ can be easily repeated for
$\eta_{1}(t,\omega_{1})$ and $\eta_{2}(t,\omega_{2}),$ which show the same
qualitative properties described above, but with different values for the
position and the width of the peak: namely, the distributions $\eta
_{1}(t,\omega_{1})$ and $\eta_{2}(t,\omega_{2})$ are peaked around the values
$\bar{\omega}_{1}$ and $\bar{\omega}_{2}$ defined in Eq. (\ref{o12b}) and have
a time-dependent width at mid height given by the rescaled quantities%
\begin{equation}
\delta\omega_{1}=\frac{\Gamma_{1}}{\Gamma}\delta\omega\text{ , }\delta
\omega_{2}=\frac{\Gamma_{2}}{\Gamma}\delta\omega\text{ ,}%
\end{equation}
where $\Gamma_{1}$ and $\Gamma_{2}$ have been defined in Eq. (\ref{gamma12})
and $\delta\omega$ is the time-dependent function defined in Eq.
(\ref{deltaomega}) and plotted in Fig. 3. Then, for each particle an enhanced
spread of the energy takes place for short times. For long times, being
$\delta\omega\rightarrow\Gamma$, the limits $\delta\omega_{1}\rightarrow
\Gamma_{1}$ and $\delta\omega_{2}\rightarrow\Gamma_{2}$ are realized. The
integrals over all the energies read:
\begin{equation}
\int_{0}^{\infty}\eta_{1}(t,\omega_{1})d\omega_{1}=\int_{0}^{\infty}\eta
_{2}(t,\omega_{2})d\omega_{2}=1-p(t)=1-e^{-\Gamma t}\text{ ;}%
\end{equation}
the overall probability to find the particle 1 (or 2) is, as it must, the
overall decay probability.

We now turn to some examples from particle and atomic physics:

(i) $\pi^{0}$ decay: the electromagnetic decay of the $\pi^{0}$ meson into two
photons, $\pi^{0}\rightarrow\gamma\gamma,$ has a mean life time of $\tau
_{\pi^{0}}=(8.52\pm0.18)\cdot10^{-17}$ s \cite{PDG}. Here $m_{1}=m_{2}=0$ and
the previous formulae simplify. Namely, each photon has an energy of
$E_{\gamma}=\omega_{1}=\omega_{2}$, whose distribution is peaked for
$\bar{\omega}_{1}=\bar{\omega}_{2}=M_{\pi^{0}}/2$, and, in virtue of the
spreading described above, the energy uncertainty per photon is given by
$\delta E_{\gamma}=\delta\omega/2$ with $\delta\omega$ given in Eq.
(\ref{deltaomega}). Thus, $\delta E_{\gamma}$ is larger than $\Gamma_{\pi^{0}%
}/2$ (where $\Gamma_{\pi^{0}}=1/\tau_{\pi^{0}}$) if a measurement at a time of
$t\lesssim3\tau_{\pi^{0}}$ is performed. This is however difficult because of
the very short times involved.

(ii) $\pi^{+}$ decay: the decay $\pi^{+}\rightarrow\mu^{+}\nu_{\mu}$ is a weak
two-body decay process with a lifetime $\tau_{\pi^{+}}=(2.6033\pm
0.0005)\cdot10^{-8}$ s. Plugging in the nominal masses (neglecting neutrino
mixing and masses) implies that $\delta E_{\mu}=\delta\omega_{1}%
\simeq0.213\delta\omega,$ thus the energy spread of the muon energy is smaller
than $\delta\omega/2$ because of the sizable muon mass. On the other hand, for
the energy uncertainty of the neutrino one has $\delta E_{\nu}=\delta
\omega_{2}\simeq0.787\delta\omega,$ thus larger than $\delta\omega/2$. While
the neutrino can hardly be detected, a measurement of the muon at such an
early time could be feasible.

(iii) Atomic spontaneous emission: an atom in which an electron is in an
excited state decays to its ground state by emitting a photon. This is indeed
the original framework in which the decay was studied \cite{ww,breit} (see
also e.g. Refs. \cite{scully,facchispont,ford,kofman} and refs. therein) and
can be seen as a particular limit of our formalism. Let $\Delta E$ be the
energy difference of the two energy levels: the initial `mother' state is the
excited atom, with a central mass $M=m_{M}=m_{D}+\Delta E$, and the final
state consists of two particles, the `daughter' state (ground-state stable
atom, with mass $m_{D}$) and a photon. Due to the fact that $\Delta E\ll
m_{D},$ it follows that $\delta E_{D}=\delta\omega_{1}\simeq0$ and $\delta
E_{\gamma}=\delta\omega_{2}\simeq\delta\omega$ (valid to a very good level of
approximation). In this case the uncertainty on the energy of the photon is
the whole energy uncertainty of the quantum system. Thus, the distribution
energy of the photon is given by $\eta_{2}(t,\omega_{2})=\eta\left(
t,\omega-m_{D}\right)  $, which is centered on $\Delta E$ and has for large
times a spread of $\delta E_{\gamma}=\delta\omega=\Gamma$, where $\Gamma$ is
the decay width of the spontaneous emission process: this energy uncertainty
is nothing else than the well-known natural broadening of the spectral line
due to the uncertainty principle. For decreasing $t$ the quantity $\delta
E_{\gamma}$ shows the broadening expressed in Fig. 3. Thus, if it were
possible to perform a measurement at a time scale of the order of $\Gamma
^{-1}$ (typically in the range of $10^{-9}$ s, but dependent on the considered
atom and energy levels), then the broadening of $\delta\omega$ could be
eventually visible. (Indeed, another known effect of spectral line broadening
is the so-called impact pressure broadening, in which other particles
interrupts the decay process. A question is if such pressure broadening can be
related to the effect described in this work: to which extent can the other
particles, and thus the environment, make a measure of the unstable excited
atom? This surely interesting topic is left as an outlook).%

\begin{figure}
[ptb]
\begin{center}
\includegraphics[
height=6.832in,
width=5.0298in
]%
{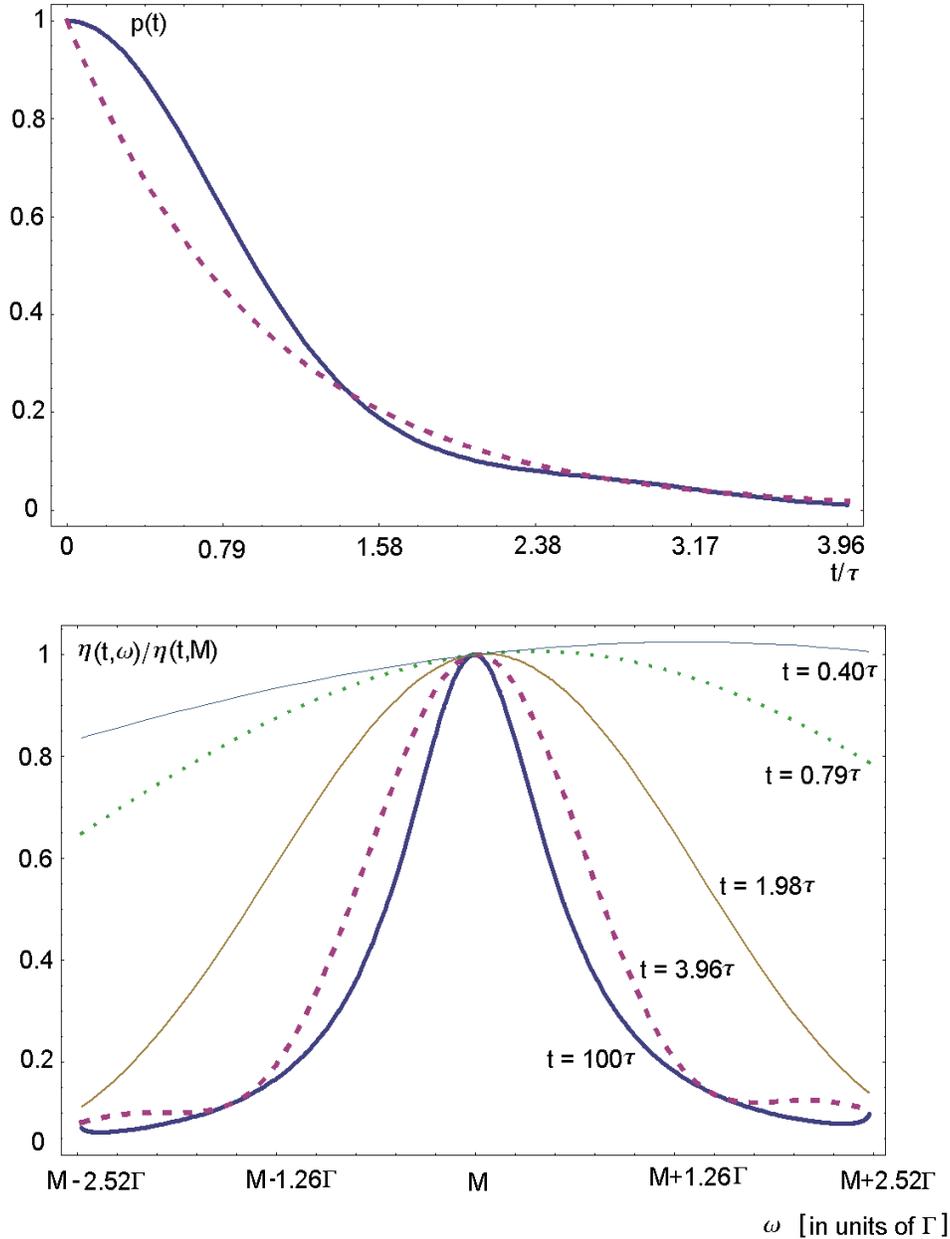}%
\caption{(Color online) Upper panel: survival probability $p(t)$ (solid line)
for the case of Eq. (\ref{fk}) and for the numerical values $g=0.95\sqrt
{\Gamma}$, $M-E_{0}=2.52\Gamma$, $\alpha=0.0396\tau$. The deviation from the
exponential form $e^{-\Gamma t}$ (dashed line) at short times is evident.
Lower panel: the function $\eta(t,\omega)/\eta(t,M)$ is plotted for different
vales of $t$. Besides the broadening, the asymmetry generated by the parameter
$\alpha$ in\ Eq. (\ref{fk}), which is hardly visible at long times (solid
thick line, $t=100\tau$), increases for decreasing $t.$ For $t=0.79\tau$ and
for $t=0.40\tau$ it is clearly visible.}%
\end{center}
\end{figure}

As a last step we go beyond the exponential limit by studying a case in which
the function $f(k)$ takes the form%
\begin{equation}
f^{2}(k)=\left(  1+\alpha k\right)  \theta(E_{0}^{2}-(E-M)^{2}). \label{fk}%
\end{equation}
For simplicity, we still keep the relation $\omega(k)=k$ valid. The step
function in Eq. (\ref{fk}) implies that the state $\left\vert S\right\rangle $
couples to a continuum of states $\left\vert k\right\rangle $ limited to a
band of energy $(M-E_{0},M+E_{0})$ \cite{gp}. In virtue of the introduced
thresholds the survival probability $p(t)$ is not an exponential for short and
long times. Moreover, the parameter $\alpha,$ which has the dimension
$[E^{-1}]$, introduces an asymmetry in the coupling to this band. As a
consequence, the spectral function $d_{S}(x)$ is not symmetric around the peak
located at $M$. The self-energy contribution takes the form (after reabsorbing
an inessential constant) $\Pi(E)=\frac{g^{2}}{2\pi}(1+\alpha E)\log\left[
\frac{E-M-E_{0}}{E-M+E_{0}}\right]  $. Note, in the limit $\alpha=0$ and
$E_{0}\rightarrow\infty$ we recover the self-energy in the exponential case:
$\Pi(E)=ig^{2}/2$.

The expression of the decay width as given by the Fermi golden rule reads now
$\Gamma=g^{2}(1+\alpha M)$ and the corresponding mean life time is
$\tau=1/\Gamma$. In this case, being the decay not exponential, $\Gamma$ is
only an approximate quantity. This is clearly visible in Fig. 4, upper panel,
where the survival probability $p(t)$ is plotted for a suitable numerical
choice ($g=0.95\sqrt{\Gamma},$ $M-E_{0}=2.52\Gamma,$ $\alpha=0.0396\tau$) and
compared to the exponential decay $e^{-\Gamma t}.$ The typical quadratic
behavior of $p(t)$ is realized; then, after some oscillations, the exponential
limit is reached (deviations for large times take place as well, but are not
relevant here).

For the purposes of our work, we study the function $\eta(t,\omega)$ for this
system for different times. Its analytic expression can be derived by the
previously presented general formulae and reads%
\begin{equation}
\eta(t,\omega)=\frac{\operatorname{Im}\Pi(\omega)}{\pi}\left\vert
\int_{-\infty}^{\infty}\mathrm{dE}d_{S}(E)\frac{e^{-i\omega t}-e^{-iEt}%
}{\omega-E}\right\vert ^{2}\text{.}%
\end{equation}
(This expression is formally valid for each choice of $f(k),$ as long as
$\omega(k)=k$. In the case of Eq. (\ref{fk}): $\operatorname{Im}\Pi
(\omega)=g^{2}\left(  1+\alpha\omega\right)  /2$ for $\omega\in$
$(M-E_{0},M+E_{0})$ .) The function $\eta(t,\omega)/\eta(t,M)$ is plotted in
Fig. 4, lower panel, for different values of time. For large times,
$t=100\tau,$ the equality $\eta(t\rightarrow\infty,\omega)=d_{S}(\omega)$ is
realized (thick solid line). It is then visible that the function
$d_{S}(\omega)$ has the usual form and that the asymmetry due to the parameter
$\alpha$ introduced in Eq. (\ref{fk}) is hardly noticeable. However, when
going to smaller times, the expected broadening takes place, thus showing the
generality of this effect. On top of that, an interesting phenomenon emerges:
the asymmetry gets enhanced. For $t=0.40\tau$ (thin solid line) the function
$\eta(t=0.40\tau,\omega)$ this is evident: even the maximum does not take
place at $M$ but is shifted to a higher value.

One important remark about the model of Eq. (\ref{fk}) is necessary: two
additional discrete energy levels emerge for each value of the coupling
constant $g$ (for small $g,$ one emergent stable state has an energy just
below $M-E_{0}$ and the other just above $M+E_{0}$.) As a consequence, the
survival probability $p(t)$ shows in general for large times oscillations
which involve these additional discrete levels, see also Ref. \cite{kofman}
for the description of this phenomenon. However, for the numerical values used
in Fig. 4 the presence of these two additional stable states is negligible:
for very large $t$ the survival amplitude reads $a(t)\simeq Z_{1}e^{-iE_{1}%
t}+Z_{2}e^{-iE_{2}t},$ where $Z_{1}=0.92\cdot10^{-6}$ and $Z_{2}%
=1.4\cdot10^{-5}$ are very small numbers, meaning that the oscillations are
extremely suppressed (and, moreover, have practically no influence on the
short-time behavior of the system); the energies $E_{1}=M-E_{0}-1.3\cdot
10^{-7}\Gamma$ and $E_{2}=M+E_{0}+2.5\cdot10^{-6}\Gamma$ are the emergent
discrete levels (very close the the energy thresholds of Eq. (\ref{fk})). In
conclusion, for the illustrative purposes of our analysis the use of the
simple model of Eq. (\ref{fk}) is acceptable, although the subtlety of the
emerging stable states should be kept in mind when this model is studied; we
also refer to Ref. \cite{wolkanowski}, where this issue has been analyzed both
analytically and numerically in great detail in dependence of the coupling
constant $g$.

The emergence of discrete energy levels for each value of $g$ is indeed a
peculiarity of Eq. (\ref{fk}) due to the sharp boundaries. A more realistic
form of $f^{2}(k)$ is given by
\begin{equation}
f^{2}(k)=(1+\alpha k)\frac{\sqrt{k-(M-E_{0})}}{k^{2}+\Lambda^{2}}%
\theta(k-(M-E_{0}))\text{ ,} \label{fknew}%
\end{equation}
in which a phase space factor $\sqrt{k-(M_{0}-E_{0})}$ renders the function
continuous close to the left energy threshold $(M-E_{0})$ and a smooth cutoff
behavior for large $k$ replaces the right threshold. For this form, no
additional stable state emerges as long as $g$ does not exceed a critical
value. (Quite interestingly, the emergence of an additional stable state when
the coupling constant exceeds a certain value takes place also in relativistic
quantum field theory, see Ref. \cite{e38}.) A numerical study of the system
with Eq. (\ref{fknew}) shows the same qualitative features as Fig. 4. Then, we
are led to think that the described properties (broadening and distortion for
small times) are rather general, and hold also for more complicated choices of
the function $f(k)$ and $\omega(k)$.

In conclusions, we have studied the energy uncertainty of the final state of a
decay process, finding that it can be much larger than the natural decay width
if a measurement of the final state and its energy is performed at time
comparable to (or smaller than) the lifetime of the unstable state, see Figs.
1, 2, and 3. Further outlooks are listed in the following. (i) Systematic
study for a class of models with short- and long-time deviations from the
exponential decay law. The simple case studied here leading to Fig. 4 shows
that this subject is potentially very interesting. (ii) Study of the energy
spread of the final state of decay processes in which the initial unstable
state can decay, along the line of Ref. \cite{duecan}, in more than one decay
channel. (iii) In this work the measurement has been still considered of the
ideal type; a description of the measurement in a more realistic way, along
the line of Ref. \cite{shimizu}, is an important work for the future. In this
framework one has to make particular attention to the details of the type of
measurement performed.

\bigskip

\textbf{Acknowledgments: } the author thanks G. Pagliara and T. Wolkanowski
for useful discussions and the Foundation of the Polytechnical Society of
Frankfurt for support through an Educator fellowship.

\end{document}